\documentclass[aps,prl,superscriptaddress,twocolumn,showpacs,showkeys,floatfix]{revtex4-1}
\usepackage{amsmath}
\usepackage{bm}
\usepackage{amssymb}
\usepackage{graphicx}
\usepackage{epstopdf}
\usepackage{epsfig}
\usepackage{color}
\usepackage{hyperref}
\usepackage{float}
\usepackage{dcolumn}
\begin{document}
\title{$PT$-Symmetric Plasmonic Metamaterials}
\author{Hadiseh Alaeian}
\affiliation{Department of Electrical Engineering, Stanford
University, Stanford, California 94305, USA}
\affiliation{Department of Materials Science and Engineering,
Stanford University, Stanford, California 94305, USA}

\author{Jennifer A. Dionne}
\affiliation{Department of Materials Science and Engineering,
Stanford University, Stanford, California 94305, USA}

\begin{abstract}
We theoretically investigate the optical properties of parity-time
(PT) symmetric three-dimensional metamaterials composed of
strongly-coupled planar plasmonic waveguides. By tuning the
loss-gain balance, we show how the initially isotropic material
becomes both asymmetric and unidirectional. Investigation of the
band structure near the material's exceptional point reveals
several intriguing optical properties, including double negative
refraction, Bloch power oscillations, unidirectional invisibility,
and reflection and transmission coefficients that are
simultaneously equal to or greater than unity. The highly tunable
optical dispersion of $PT$-symmetric metamaterials provides a
foundation for designing an entirely new class of three-dimensional
bulk synthetic media, with applications ranging from lossless
sub-diffraction-limited optical lenses to non-reciprocal
nanophotonic devices.
\end{abstract}


\maketitle

Textbook conceptions of light-matter interactions have been
challenged by two recent material advances --- the development of
metamaterials and the discovery of parity-time $PT$-symmetric
media. Metamaterials allow considerable control over the electric
and magnetic fields of light, so that permittivities,
permeabilities, and refractive indices can be tuned throughout
positive, negative, and near-zero values. Metamaterials have
enabled negative refraction, optical lensing below the diffraction
limit of light and invisibility cloaking~\cite{Valentine08,Yao08,
Zhang08, Zhang09, Atre13, Xu13, Shalaev07, Soukoulis11, Lezec07}.
Complementarily, $PT$-symmetric media allow control over
electromagnetic field distributions in loss and gain media, so that
light propagation can be asymmetric and even unidirectional.
$PT$-symmetric media have enabled loss-induced optical
transparency, lossless Talbot revivals and unidirectional
invisibility~\cite{Benisty11, Guo09, Ramezani12, Makris10, Zheng10,
Longhi11, Ruter10, Lin11, Mostafazadeh13, Feng13, Makris08,
Castald13}. Combined with non-linear media, they have also been
suggested as optical diodes, insulators, circulators, and perfect
cavity absorber-lasers~\cite{Musslimani08, Lazarides13, Li12, He11,
Dmitriev10, Chong11, Longhi10}.

While metamaterials rely on subwavelength engineered `building
blocks' to control electric and magnetic light-matter interactions,
$PT$-symmetric media rely on judicious spatial arrangement of loss
and gain media. Their unique asymmetric properties are based on a
fundamental insight from quantum mechanics indicating that
Hamiltonians need not be Hermitian to yield real eigenvalues and
hence physical observables. Instead, the weaker condition of parity
and time symmetry is sufficient to yield real eigenvalues below a
certain threshold. Above this threshold, eigenvalues move into the
complex plane and become complex conjugates of each
other~\cite{Bender99, Bender07, Ahmed01, Bender98, Mostafazadeh02}.

In the context of optics, $PT$-symmetric Hamiltonians arise from
the duality between the quantum mechanical Schrodinger equation and
the wave equation. Provided the refractive index profile satisfies
$n(x)=n*(-x)$, light will propagate as if it experiences a
$PT$-symmetric potential. Below the $PT$-symmetric exceptional
point, the optical eigenvalues will be purely real; however, as the
loss and gain of the material are increased beyond the exceptional
point, the eigenvalues will become complex. In particular, certain
eigenmodes will experience increased loss while other eigenmodes
will exhibit strong optical gain. This behavior is at the core of
the asymmetric and unidirectional optical properties observed in
$PT$ media to date.

While nearly all $PT$-symmetric media have been constructed from
macroscopic (i.e., greater than wavelength-scale) elements, the
optical Hamiltonian places no restrictions on the length scales
over which the index profile can vary. This insight drives the
question: can we create $PT$-symmetric metamaterials --- i.e., bulk
photonic media whose optical properties are determined both by
their subwavelength building blocks and a judicious choice of their
loss/gain profile? Such metamaterials would enable unprecedented
control over electric and magnetic optical fields across wavelength
and subwavelength scales, and may enable an entirely new class of
bulk synthetic photonic media.

In this Letter, we investigate the emergent optical properties of
bulk, three-dimensional $PT$-symmetric metamaterials. As a
prototype metamaterial, we consider a multilayer stack of
alternating layers of metal and dielectric. Both
theoretical~\cite{Verhagen10} and experimental~\cite{Xu13} work has
demonstrated the isotropic negative index response of this
metamaterial, resulting in all-angle negative refraction and
Veselago `perfect optical lensing.' Its operation is based on the
negative index plasmonic modes of its unit cell --- a five-layer
`metal-insulator-metal' waveguide. By varying the thickness of the
layers as well as the materials, the frequency of operation and the
emergent bulk index of refraction can be precisely controlled
throughout optical frequencies. While practical utilization of this
negative index metamaterial has been limited by propagation and
coupling losses, we will show that these losses could be overcome
by subjecting the plasmonic modes to $PT$-symmetric potentials.
Moreover, $PT$-symmetric potentials in this metamaterial can enable
above-unity transmission and reflection, Bloch power oscillations,
hyperbolic to elliptic dispersion transitions, and unidirectional
invisibility.

\begin{figure}
\includegraphics[scale=1]{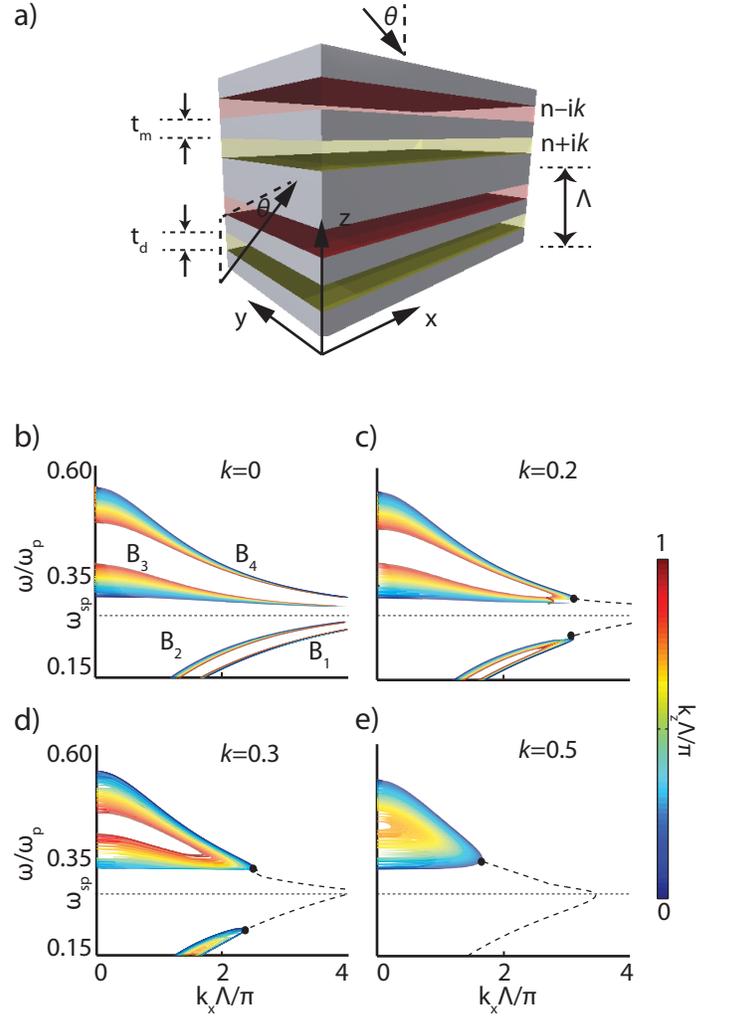}
\caption{\label{fig1} Schematic of the plasmonic metamaterial (a), and
the associated band diagrams for transverse magnetic modes (TM). The
metamaterial is composed of coupled plasmonic waveguides, with
deeply subwavelength metal and dielectric thicknesses of
$t_m$=$t_d$= 30 nm. The unit-cell period is $\Lambda$ = 150 nm. The
refractive index of the dielectric is $n \pm ik$, with $n$ = 3.2 and
\emph{k} corresponding to the non-Hermiticity parameter. Band
diagrams are included for \emph{k}= b)0, c)0.2, d)0.3 and e)0.5.
Bands $B_1$ and $B_2$ are positive refractive index bands, while
bands $B_3$ and $B_4$ correspond to negative refractive index
bands. The horizontal dotted black lines in panels (b)-(e) indicate
the surface plasmon resonance frequency, $\omega_{sp}$. The black
circles in panels in the same panels indicate the exceptional points of the
dispersion, beyond which the purely real modes evolve to the complex
modes characterized by either loss or gain (dashed black lines).}
\end{figure}

Figure~\ref{fig1}a illustrates the specific plasmonic metamaterial
investigated in this paper, with the unit cell period indicated by
$\Lambda$. Within each unit cell, the thicknesses of the metal
$t_m$ and dielectric $t_d$ are deeply subwavelength, with $t_m$ =
$t_d$ = 30 ~nm. We consider Ag as the metal, described by a
lossless Drude model with dielectric constant
$\epsilon_{Ag}=1-(\frac{\omega_p}{\omega})^2$ \footnote{Here we
assume a lossless Drude model, but the results are generalizable to
a realistic case including loss. See for example reference
~\cite{Benisty11}.} The bulk plasma frequency of Ag, $\omega_p$, is
assumed to be $8.85\times 10^{15}$s$^{-1}$. We consider the
dielectric layers to be TiO$_2$ with n = 3.2. With these materials,
the surface plasmon resonance, $\omega_{sp}$, occurs at 1.73 eV
$(\omega_{sp}/\omega_p=0.29)$, and negative index modes are
observed between this frequency and $\omega_p$. This particular
materials combination was recently experimentally shown to exhibit
all-angle negative refraction and Veselago lensing~\cite{Xu13}.
Here, we theoretically investigate the evolution of the optical
bands of this metamaterial upon varying the imaginary part of the
refractive index of TiO$_2$, denoted as \emph{k} in Fig.\ref{fig1}.
$PT$-symmetric potentials require balanced loss and gain, so the
magnitude of this `non-Hermiticity parameter', \emph{k}, is
identical for alternating dielectric layers.

Using the transfer matrix approach described in~\cite{Russell95},
we solve for the dispersion curves of the five-layer unit-cell
plasmonic waveguide for transverse-magnetic (TM) polarized
illumination. To determine the band diagrams of the periodic
metamaterial, the wavevector along the \emph{z}-direction is swept
in the first Brillouin zone, $(0,\frac{\pi}{\Lambda})$, and the
characteristic equation is minimized to find the propagation
constant along the \emph{x}-direction at each frequency. The
results are shown in panels (b)-(e) of Fig.~\ref{fig1} for \emph{k}
= 0, 0.2, 0.3 and 0.5, respectively. Note that the colormap
indicates purely real values of $k_x$, corresponding to lossless
propagation along the metamaterial. For a non-Hermiticity parameter
\emph{k} = 0, four different branches are observed: two below
$\omega_{sp}$ (B$_1$ and B$_2$) and two above (B$_3$ and B$_4$).
Because all constituents are lossless, the wavevectors diverge at
$\omega_{sp}$. B$_1$ and B$_2$ are characterized by positive slopes
and hence positive refractive mode indices. In contrast, B$_3$ and
B$_4$ are characterized by negative slopes and hence negative
refractive mode indices.

When the non-Hermiticity parameter of the metamaterial is
increased, the modes merge together at the exceptional points of
the dispersion, denoted by black circles in panels (c)-(e). Beyond
these exceptional points, the two distinguishable lossless modes
below and above $\omega_{sp}$ (i.e., B$_1$ and B$_2$ or B$_3$ and
B$_4$, respectively) evolve to a gain mode and a
loss mode with the same phase velocity. Due to their complex
wavevectors, we denote these modes as black, dashed lines in
Fig.~\ref{fig1}(c)-(e). To understand these loss and gain modes,
note that the transfer matrix of the $PT$-symmetric metamaterial
possesses the following symmetry property:
\begin{equation} \label{eq: T-matrix1}
T(\omega,k_z,k_x^*)T^*(\omega,k_z,k_x)=I
\end{equation}
where \emph{I} is the identity matrix. The Bloch modes of the
metamaterial are eigenvalues of \emph{T} and satisfy:
\begin{equation}\label{eq: ch1}
|T(k_x)-e^{i\Lambda k_z}I|=0
\end{equation}
Taking the complex conjugate of Eq.~\ref{eq: ch1} and using the
symmetry property of Eq.~\ref{eq: T-matrix1}, the following
relation is obtained:
\begin{equation}\label{eq: ch2}
|T(k_x^*)-e^{i\Lambda k_z}I|=0
\end{equation}
Equation ~\ref{eq: ch2} means that if $k_x$ (a complex number in
general) admits a real solution for the Bloch wavevector, $k_x^*$
is a solution for that Bloch mode as well. Accordingly, the bands
have centro-symmetry in the complex $(k_x,k_z)$ plane. Also note that
the loss and gain modes of Fig.~\ref{fig1} (c)-(e) conjoin at
$\omega_{sp}$; further, unlike the modes for a zero non-Hermiticity
parameter, their wavevectors at $\omega_{sp}$ remain finite.

While \emph{real} periodic spatial refractive index profiles lead
to the appearance of an infinite number of band gaps,
\emph{complex} periodic index profiles generally result in complex
dispersion curves across the entire frequency range. Interestingly,
if the refractive index profile satisfies the condition for $PT$
symmetry ($n(z)=n^*(-z)$) real propagation constants and complete
band gaps can exist provided $k \leq k_{th}$. Here, $k_{th}$ is the
threshold value at which the Hamiltonian and the $PT$ operator no
longer commute, and consequently, real-valued solutions cease to be
supported by the complex potential. Fig.~\ref{fig1} (c)-(e)
illustrate this feature for increasing non-Hermiticity parameter.
For example, for \emph{k} = 0.2 and \emph{k} = 0.3 purely real
wavevectors and bandgaps are observed for all bands both above and
below $\omega_{sp}$. However, for \emph{k} = 0.5, purely real
eigenmodes below $\omega_{sp}$ do not exist across visible and
near-infrared frequencies. Further, the bandgap between $B_3$ and
$B_4$ merges for large $k_z$, and these bands only exist over a
very limited wavevector and wavelength range.

\begin{figure}
\includegraphics[scale=0.9]{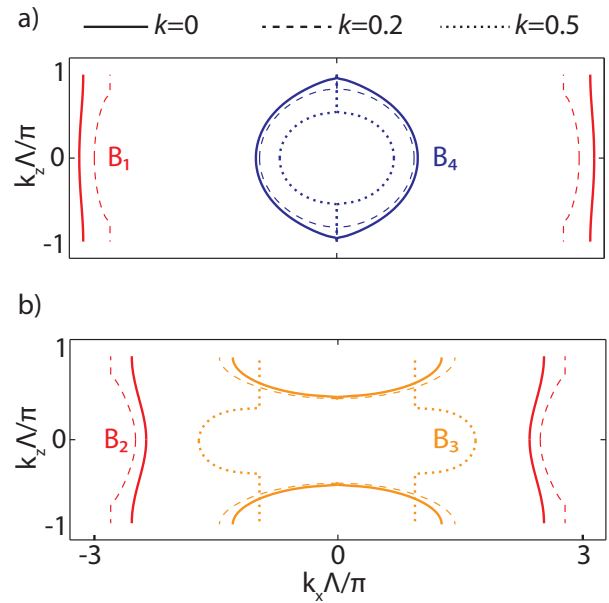}
\caption{\label{fig2} Equi-frequency contours of the metamaterial,
plotted for both positive and negative index bands at wavelengths
of $\lambda$ = 954 nm ($B_1$,$B_2$), 604 nm ($B_3$) and 445 nm
($B_4$). Band $B_4$ is circular for zero and small $k$,
corresponding to an isotropic metamaterial. For all
$k$, bands $B_1$ and $B_4$ remain elliptical while band $B_2$ is
hyperbolic. However, band $B_3$ undergoes a hyperbolic to
elliptical transition with increasing $k$.}
\end{figure}

The non-Hermiticity parameter not only changes the propagation
constant and bandgap of the metamaterial, but also the band
curvature. Fig.~\ref{fig2} plots the equi-frequency contours of
bands $B_1$-$B_4$ at wavelengths of $\lambda=954$ nm
($\omega/\omega_p=0.22$) for $B_1$ and $B_2$, $\lambda=604$ nm
($\omega/\omega_p=0.35$) for $B_3$, and $\lambda=445$ nm
($\omega/\omega_p=0.48$) for $B_4$. Since the metamaterial is
isotropic in the \emph{xy}-plane and the contours are
centro-symmetric in the $(k_z,k_x)$ plane, a quadratic dispersion
relation $(\frac{k_x}{n_x})^2+(\frac{k_z}{n_z})^2=k_0^2$ can be
used to model the bands. Here $k_0$ indicates the free-space
wavevector. The fitted refractive mode indices are listed in
Table~\ref{table1}.

\begin{table}[h!]
\begin{center}
\begin{tabular}{| l | l | l | l |}
\hline
& \emph{k} = 0 & \emph{k} = 0.2 & \emph{k} = 0.5 \\ \hline
& $(n_x^2$ , $n_z^2)$ & $(n_x^2$ , $n_z^2)$ & $(n_x^2$ , $n_z^2)$ \\ \hline
$B_1$ & (95.86 , 289.89) & (85.5 , 53.56) & NA \\ \hline
$B_2$ & (54.14 , $-$58.86) & (59.32 , $-$24.83) & NA \\ \hline
$B_3$ & ($-$4.11 , 1.27)   & ($-$4.88 , 1.21) & (11.55 , 1.16) \\ \hline
$B_4$ & (1.9 , 1.85) & (1.85 , 1.45) & (1.02 , 0.66) \\ \hline

\end{tabular}
\end{center}
\caption{\label{table1} Effective refractive indices of the four bands based on a
quadratic fit. The parameters are calculated for wavelengths of
$\lambda$ = 954 nm ($B_1$,$B_2$), 604 nm ($B_3$) and 445 nm ($B_4$).}
\end{table}

As seen both in Fig.~\ref{fig2} and Table~\ref{table1}, for a non-Hermiticity parameter \emph{k} = 0, bands
$B_1$ and $B_4$ are elliptical (i.e., $n_x^2n_z^2 \geq$0) while
bands $B_2$ and $B_3$ are hyperbolic (i.e., $n_x^2n_z^2 \leq$0).
Moreover, $B_4$ is characterized by a nearly-perfect circular
equi-frequency contour and almost equal values of effective
refractive indices in both the \emph{x} and \emph{z}-directions.
Accordingly, this metal-insulator-metal metamaterial is isotropic
at $\lambda$ = 445 nm, consistent with prior work~\cite{Verhagen10, Xu13}.

For increasing non-Hermiticity parameter, $B_1$ and $B_4$ remain
elliptical while $B_2$ remains hyperbolic. Interestingly, however,
band $B_3$ undergoes a hyperbolic to elliptical transition for
\emph{k} = 0.5. Such hyperbolic-to-elliptic transitions could enable
dynamic tuning of Purcell enhancements for emitters near the
metamaterial. Further, they could modulate Talbot revivals or the
formation and resolution of images generated by hyperbolic
metamaterial super-lenses~\cite{Ni11, Jacob12, Webb06, Kim12,Zhao11}.

The results of Fig.~\ref{fig2} imply that with increasing
non-Hermiticity parameter, the material can evolve from an
isotropic metamaterial to an anisotropic one. Intriguingly, the
structure can also become highly directional. This property cannot
be derived from the band diagrams, but can be understood by
considering the transfer matrix:
\begin{equation}
T=
\begin{bmatrix}
a & b \\ c & a^*
\end{bmatrix}
\end{equation}
Here, the parameters $a$, $b$ and $c$ are related to the reflection
and transmission coefficients $r$ and $t$ as $r_L=-\frac{c}{a^*}$,
$r_R=\frac{b}{a^*}$ and $t_L=t_R=\frac{1}{a^*}$, where the
subscripts L and R denote illumination from the left and right,
respectively. As these equations indicate, an optical system
composed of linear and reciprocal materials is non-directional
provided the components are lossless. In other words, the
transmitted and reflected powers $T=|t|^2$ and $R=|r|^2$ sum to
unity and are independent of illumination direction, since
$T_L=T_R=T$ and $T+R_R=1=T+R_L$, so $R_L=R_R$. When loss or gain is
introduced into the system, the transmission coefficient remains
the same for both directions of illumination. However the
reflection coefficient need not be symmetric, as power can be
attenuated or generated within the structure. The asymmetry is
obtained at the price of losing propagating Bloch modes. However,
as we will show, asymmetric responses can be obtained in a
$PT$-symmetric potential where purely real bands exist as well.

\begin{figure}
\includegraphics[scale=1]{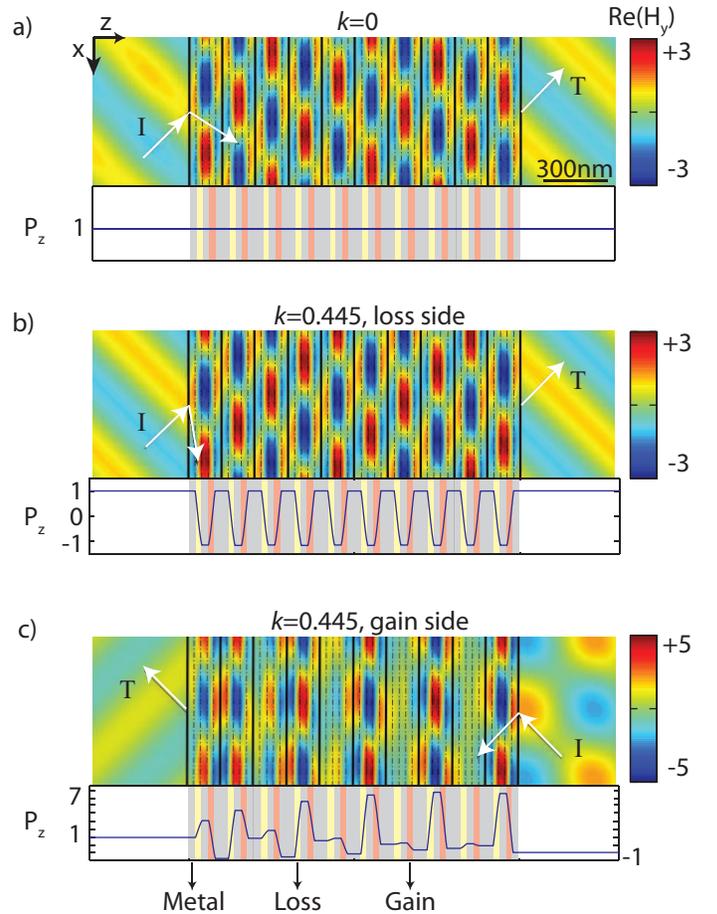}
\caption{\label{fig3} Real($H_y$) for a 10-unit-cell metamaterial
upon illumination by a planewave in air of wavelength $\lambda$ =
445 nm and angle $\theta$ = 45$^\circ$. Panel (a) shows the results
for \emph{k} = 0 while (b) and (c) show the results for \emph{k} =
0.445 when illuminated from loss ($+z$) and gain ($-z$) side,
respectively. The overlaid arrows indicate the direction of the
averaged Poynting vectors in the corresponding regions. I in each
panel indicates the \emph{incident} wavevector while T shows the
\emph{transmitted} planewave. The lower graphs in each panel show
the distribution of the power in each layer of the stack. The
metallic layer is represented in gray, while the gain and loss
regions are represented in red and yellow, respectively.}
\end{figure}

To illustrate this directional behavior, Fig.~\ref{fig3} plots
plane-wave refraction of light from vacuum ($n=1$) through a
metamaterial composed of 10 unit cells. We consider TM-polarized
illumination of wavelength $\lambda$ = 445 nm impinging on the
metamaterial at an angle of $\theta$ = 45$^\circ$ in the $(x,z)$
plane. The colormap of Fig.~\ref{fig3} plots the $H_y$ component of
the fields. The arrows of Fig.~\ref{fig3} indicate the direction of
illumination, refraction and transmission, each determined by
spatially-averaging the Poynting vector in each region.

\begin{figure}
\includegraphics[scale=0.95]{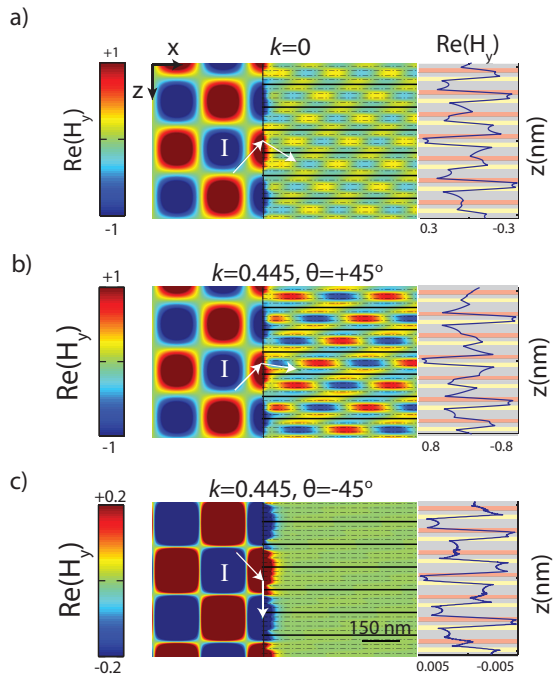}
\caption{\label{fig4} Real($H_y$) when a semi-infinite metamaterial
is illuminated with a planewave at $\lambda$ = 445 nm along
\emph{x}-direction (endfire illumination). The illumination angle
$\theta$ equals 45$^\circ$ in (a) and (b) and $-45^\circ$ in (c).
The side graphs in each panel show the distribution of the magnetic
field 1$\mu$m away from the interface.}
\end{figure}

For \emph{k} = 0 (Fig.~\ref{fig3}a), the power is negatively
refracted with an angle of $\sim-32^\circ$. This result is in
excellent agreement with our band structure calculations, which
yield a refracted angle from Snell's law of $\sim-31^\circ$. The
refractive index, $n=-\sqrt{1.87}$ = $-$1.36, at this
non-Hermiticity value is independent of the illumination angle and
direction. Indeed, for illumination in the $(x,z)$ plane, or an
`endfire configuration', the same refraction angle is observed (see
Fig.~\ref{fig4}a). The same refraction angle is also observed for
illumination from all sides of the metamaterial (i.e., illumination
from $\pm x$, $\pm y$, and $\pm z$)

Upon increasing the non-Hermiticity parameter of the metamaterial,
the material becomes highly directional. Panels (b) and(c) of
Fig.~\ref{fig3} illustrate the field profiles in a 10-layer
$PT$-symmetric metamaterial when illumination is from the loss and
gain side (i.e., illumination in the $+z$ or $-z$ directions,
respectively). As a particular example, we consider \emph{k} =
0.445. As seen, field profiles and refraction angles are completely
different for illumination from +$z$ (loss side) versus $-z$ (gain
side). Illumination from +$z$ yields negative refraction at an
angle of $~-81^\circ$ (see Fig.~\ref{fig3}b). In contrast,
illumination from $-z$ yields negative refraction at an angle of
$\sim-43^\circ$ (Fig.~\ref{fig3}c).

More intriguingly, this structure is characterized by tunable
reflection and transmission coefficients that can range from zero
to at or above unity. This characteristic is illustrated in the
lower panels of Fig.~\ref{fig3}, which plot the
normalized-to-incidence power at each position along the direction
of propagation. For example, for illumination from the $-z$
direction (panel (c)) power flows backward toward the source in the
illumination region ($P_z=-1$), and away from the metamaterial on
the transmission side ($P_z=+1$). Therefore, the metamaterial is
completely transparent, in that the metamaterial can transmit all
of the incident power, even though light is emitted back towards
the source.

Moreover, for illumination from the $+z$ direction (panel (b)) this
metamaterial can achieve unidirectional invisibility. As seen, the
power is unity on both sides of the metmaterial, indicating
complete suppression of reflection on the illumination side and
complete transmission on the other. Formally, perfect invisibility
requires that the transmitted phase ($\phi_{t}$) equal the phase of
a planewave propagating in free-space ($\phi_{FS}$). For the
10-layer metamaterial of Fig.~\ref{fig3}b,
$\frac{\phi_{FS}-\phi_t}{2\pi}$ = 2.75, so the object could be
identified through the interference with a reference planewave.
However, perfect unidirectional invisibility, i.e.
$\phi_{FS}-\phi_t=2m\pi$ where $m$ is an integer, can be achieved
when the number of periods is increased to 55, 74 and 91.

The unusual directional scattering properties of the metamaterial
can be rationalized by considering the power as light propagates
through the array. For \emph{k} = 0, power remains constant
throughout the metamaterial (Fig.\ref{fig3}a). However, with
increasing \emph{k}, power begins to oscillate within the
metamaterial, with power increasing in the gain regions and
decreasing in the loss regions. These plasmonic power oscillations
are analogous to Bloch oscillations observed in both electronic and
photonic crystals~\cite{Sapienza03} as well as PT-symmetric
arrays~\cite{Musslimani08}.

As a final visual example of the unusual unidirectional properties
of this metamaterial, Fig.~\ref{fig4} plots the fields and
refracted angles for illumination along the $+x$ direction (endfire
illumination). For a non-Hermiticity parameter \emph{k} = 0,
illumination at $\theta=\pm45^\circ$ yields refraction at
$\mp$30$^\circ$, respectively, in good agreement with the
previously determined value of $\mp$31$^\circ$. With increasing
non-Hermiticity parameter, however, illumination at $+\theta$
yields markedly different results than illumination at $-\theta$.
For example, for \emph{k} = 0.445, illumination at $+45^\circ$
yields a refracted angle of $~-11^\circ$, while illumination at
$-45^\circ$ yields refraction along the metamaterial interface at
an angle of $-90^\circ$. This double refraction does not just
manifest itself in the intensity of the transmitted beam, but also
in the profile of the fields, as seen in the right panels of
Fig.~\ref{fig4}.

We now consider the effect of varying the non-Hermiticity parameter
on the scattering properties of the metamaterial. As before, we
consider TM-polarized illumination with a 45$^\circ$ tilted
planewave at $\lambda$ = 445 nm. We limit our analysis to
illumination along either $+z$ (`left-side illumination (L)') or
$-z$ (`right-side illumination (R)') Based on Eq.~\ref{eq:
T-matrix1}, a generalized energy conservation formula can be
derived as $|T-1|=\sqrt{R_RR_L}$, where $T$ is the transmitted
power and $R_R$ and $R_L$ are the reflected powers from the right
and left sides, respectively ~\cite{Ge12}.

\begin{figure}
\includegraphics[scale=0.95]{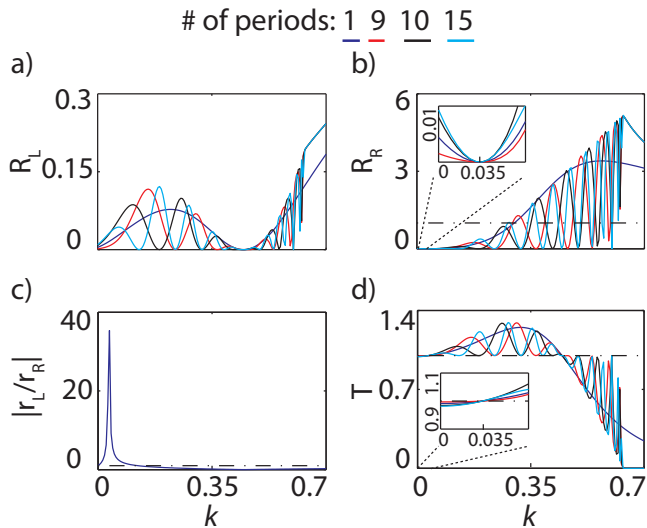}
\caption{\label{fig5} Reflected and transmitted powers of
the metamaterial with increasing $k$ and varying numbers of
periods. (a) Reflected power upon illumination from the
left (loss) side of the metamaterial; (b) reflected power upon
illumination from the gain (right) right side; (c) Quotient of
reflected powers; (d) Transmitted power. In all panels, the
incident angle is $\theta$ = 45$^\circ$ and the illumination
wavelength is $\lambda$ = 445 nm. The inset in panel (b) and (d) show the behavior of the reflected and transmitted powers around \emph{k} = 0.035 where $R_R$ vanished and $T$ = 1 independent of the number of periods.}
\end{figure}

Figure~\ref{fig5} plots the reflection and transmission
coefficients as a function of \emph{k}. As seen in
Fig.~\ref{fig5}d, for $\emph{k}$ = 0.035, the transmitted power
equals unity independent of the number of layers. Correspondingly
at this point $R_R$, shown in panel (b), vanishes for any number of
layers. This property manifests itself as a peak in
Fig.~\ref{fig4}c, where the quotient of relative reflection
coefficients is plotted. Importantly, for $\emph{k}$ = 0.035, this
$PT$-symmetric metamaterial is still isotropic, characterized by
circular equifrequency contours. Therefore, this PT-symmetric
optical potential could enable lossless and far-field Vesalago
lensing.

For larger non-Hermiticity parameters ($\emph{k}>$0.035), $T$
exceeds unity. Interestingly, while the transmitted power varies
with $\emph{k}$ and the number of unit cells, it never drops below
1 up through $\emph{k}$ = 0.445. At this non-Hermiticity value, the
reflected power from the left/loss side ($R_L$) vanishes, as shown
in Fig.~\ref{fig5}a. For larger $\emph{k}$, $T$ remains at or
below unity. Non-Hermiticity parameters $\emph{k}$ above 0.63 yield
purely imaginary $k_z$, so no propagation is allowed through the
metamaterial. This property is accompanied by a
rapid drop in $T$ and strong increase in the reflectance for any number of layer.

While for left- (or loss)-side illumination the reflection
coefficient is always less than unity, for right- (or gain)-side
illumination the reflected power can exceed 1. This behavior is not
monotonic however, depending strongly on the number of unit-cell
layers and $\emph{k}$. Indeed, only at $\emph{k}$ = 0 and
$\emph{k}$ = 0.12 are the reflected powers from the left and right
identical and equal to unity. For $\emph{k}>$0.12, the ratio
$r_L/r_R$ remains below unity (Fig.~\ref{fig5}c).

Table~\ref{table2} summarizes some of the interesting scattering
properties of the 10-unit-cell metamaterial extracted from
Fig.~\ref{fig5}. For \emph{k} = 0, the metamaterial is Hermitian
and lossless. With increasing $PT$-symmetric potentials, unusual
features such as above unity transmission (\emph{k} = 0.26), above
unity reflection (\emph{k} = 0.445) and uni-directional reflection
suppression (\emph{k} = 0.445 and \emph{k} = 0.5) are observed. The
normalized Bloch wavevector for each \emph{k} is included in the
last row of the table. Since $k_z$ is purely real, light propagates
in a pass band of the metamaterial for all of these non-Hermiticity
parameters.
\begin{table}[h]
\begin{center}
\begin{tabular}{| l | l | l | l | l | l |}
\hline
& \emph{k} = 0 & \emph{k} = 0.035 & \emph{k} = 0.26 & \emph{k} = 0.445 & \emph{k} = 0.5\\ \hline
$R_L$ & 0.0080 & 0.03 & 0.0982 & 0 & 5$\times 10^{-4}$ \\ \hline
$R_R$ & 0.0080 & 0& 0.8397 & 2 & 0.1219 \\ \hline
$T$ & 0.9920 & 1& 1.2872 & 1 & 0.9922 \\ \hline
$\frac{\Lambda k_z}{\pi}$ & 0.7633 & 0.7609 & 0.6502 & 0.4683 & 0.3948 \\ \hline
\end{tabular}
\end{center}
\caption{\label{table2} Transmitted and reflected powers from left and right and the normalized Bloch wavevector for 10-unit-cell metamaterial at $\lambda$ = 445 nm and $\theta$ = 45$^\circ$.}
\end{table}

In conclusion, we have introduced the concept of a $PT$-symmetric
metamaterial. The original lossless metamaterial, composed of a
periodically-stacked 5-layer plasmonic waveguide, was designed to
behave as an isotropic, three-dimensional negative refractive index
material. By subjecting the plasmonic modes to $PT$-symmetric
optical potentials, we demonstrated the broad tunability of the
band curvatures, band gaps and effective refractive indices of the
material. Small but non-zero non-Hermiticity parameters increased
the transmission of the isotropic negative index metamaterial to
unity. Larger non-Hermiticity parameters morphed the material from
isotropic to anisotropic and directional. The highly unusual
optical properties of $PT$-symmetric metamaterials could be used to
devise an entirely new class of bulk synthetic media, ranging from
lossless Veselago lenses to unidirectional metamaterial-based
invisibility cloaks and new non-reciprocal nanophotonic devices.

Useful discussions and feedback from Dionne Group members are
highly appreciated. This work is supported in part by a SLAC
National Accelerator Laboratory LDRD award in concert with the
Department of Energy, Office of Basic Energy Sciences, Division of
Materials Sciences and Engineering, under contract
DE-AC02-76SF00515. Funding from the Hellman Faculty Scholars
program and a National Science Foundation CAREER Award
(DMR-1151231) are also gratefully acknowledged.

\bibliographystyle{apsrev4-1}
\bibliography{reference}

\begin{thebibliography}{43}%
\makeatletter
\providecommand \@ifxundefined [1]{%
 \@ifx{#1\undefined}
}%
\providecommand \@ifnum [1]{%
 \ifnum #1\expandafter \@firstoftwo
 \else \expandafter \@secondoftwo
 \fi
}%
\providecommand \@ifx [1]{%
 \ifx #1\expandafter \@firstoftwo
 \else \expandafter \@secondoftwo
 \fi
}%
\providecommand \natexlab [1]{#1}%
\providecommand \enquote  [1]{``#1''}%
\providecommand \bibnamefont  [1]{#1}%
\providecommand \bibfnamefont [1]{#1}%
\providecommand \citenamefont [1]{#1}%
\providecommand \href@noop [0]{\@secondoftwo}%
\providecommand \href [0]{\begingroup \@sanitize@url \@href}%
\providecommand \@href[1]{\@@startlink{#1}\@@href}%
\providecommand \@@href[1]{\endgroup#1\@@endlink}%
\providecommand \@sanitize@url [0]{\catcode `\\12\catcode `\$12\catcode
  `\&12\catcode `\#12\catcode `\^12\catcode `\_12\catcode `\%12\relax}%
\providecommand \@@startlink[1]{}%
\providecommand \@@endlink[0]{}%
\providecommand \url  [0]{\begingroup\@sanitize@url \@url }%
\providecommand \@url [1]{\endgroup\@href {#1}{\urlprefix }}%
\providecommand \urlprefix  [0]{URL }%
\providecommand \Eprint [0]{\href }%
\providecommand \doibase [0]{http://dx.doi.org/}%
\providecommand \selectlanguage [0]{\@gobble}%
\providecommand \bibinfo  [0]{\@secondoftwo}%
\providecommand \bibfield  [0]{\@secondoftwo}%
\providecommand \translation [1]{[#1]}%
\providecommand \BibitemOpen [0]{}%
\providecommand \bibitemStop [0]{}%
\providecommand \bibitemNoStop [0]{.\EOS\space}%
\providecommand \EOS [0]{\spacefactor3000\relax}%
\providecommand \BibitemShut  [1]{\csname bibitem#1\endcsname}%
\let\auto@bib@innerbib\@empty
\bibitem [{\citenamefont {Valentine}\ \emph {et~al.}(2008)\citenamefont
  {Valentine}, \citenamefont {Zhang}, \citenamefont {Zentgraf}, \citenamefont
  {Avila}, \citenamefont {Genov}, \citenamefont {Bartal},\ and\ \citenamefont
  {Zhang}}]{Valentine08}%
  \BibitemOpen
  \bibfield  {author} {\bibinfo {author} {\bibfnamefont {J.}~\bibnamefont
  {Valentine}}, \bibinfo {author} {\bibfnamefont {S.}~\bibnamefont {Zhang}},
  \bibinfo {author} {\bibfnamefont {T.}~\bibnamefont {Zentgraf}}, \bibinfo
  {author} {\bibfnamefont {E.~U.}\ \bibnamefont {Avila}}, \bibinfo {author}
  {\bibfnamefont {D.~A.}\ \bibnamefont {Genov}}, \bibinfo {author}
  {\bibfnamefont {G.}~\bibnamefont {Bartal}}, \ and\ \bibinfo {author}
  {\bibfnamefont {X.}~\bibnamefont {Zhang}},\ }\href@noop {} {\bibfield
  {journal} {\bibinfo  {journal} {Nature}\ }\textbf {\bibinfo {volume} {455}},\
  \bibinfo {pages} {376} (\bibinfo {year} {2008})}\BibitemShut {NoStop}%
\bibitem [{\citenamefont {Yao}\ \emph {et~al.}(2008)\citenamefont {Yao},
  \citenamefont {Liu}, \citenamefont {Liu}, \citenamefont {Wang}, \citenamefont
  {Sun}, \citenamefont {Bartal}, \citenamefont {Stacy},\ and\ \citenamefont
  {Zhang}}]{Yao08}%
  \BibitemOpen
  \bibfield  {author} {\bibinfo {author} {\bibfnamefont {J.}~\bibnamefont
  {Yao}}, \bibinfo {author} {\bibfnamefont {Z.}~\bibnamefont {Liu}}, \bibinfo
  {author} {\bibfnamefont {Y.}~\bibnamefont {Liu}}, \bibinfo {author}
  {\bibfnamefont {Y.}~\bibnamefont {Wang}}, \bibinfo {author} {\bibfnamefont
  {C.}~\bibnamefont {Sun}}, \bibinfo {author} {\bibfnamefont {G.}~\bibnamefont
  {Bartal}}, \bibinfo {author} {\bibfnamefont {A.~M.}\ \bibnamefont {Stacy}}, \
  and\ \bibinfo {author} {\bibfnamefont {X.}~\bibnamefont {Zhang}},\
  }\href@noop {} {\bibfield  {journal} {\bibinfo  {journal} {Science}\ }\textbf
  {\bibinfo {volume} {321}},\ \bibinfo {pages} {930} (\bibinfo {year}
  {2008})}\BibitemShut {NoStop}%
\bibitem [{\citenamefont {Zhang}\ \emph {et~al.}(2008)\citenamefont {Zhang},
  \citenamefont {Genov}, \citenamefont {Wang}, \citenamefont {Liu},\ and\
  \citenamefont {Zhang}}]{Zhang08}%
  \BibitemOpen
  \bibfield  {author} {\bibinfo {author} {\bibfnamefont {S.}~\bibnamefont
  {Zhang}}, \bibinfo {author} {\bibfnamefont {D.~A.}\ \bibnamefont {Genov}},
  \bibinfo {author} {\bibfnamefont {Y.}~\bibnamefont {Wang}}, \bibinfo {author}
  {\bibfnamefont {M.}~\bibnamefont {Liu}}, \ and\ \bibinfo {author}
  {\bibfnamefont {X.}~\bibnamefont {Zhang}},\ }\href@noop {} {\bibfield
  {journal} {\bibinfo  {journal} {Phys. Rev. Lett.}\ }\textbf {\bibinfo
  {volume} {101}},\ \bibinfo {pages} {047401} (\bibinfo {year}
  {2008})}\BibitemShut {NoStop}%
\bibitem [{\citenamefont {Zhang}\ \emph {et~al.}(2009)\citenamefont {Zhang},
  \citenamefont {Park}, \citenamefont {Li}, \citenamefont {Lu}, \citenamefont
  {Zhang}, ,\ and\ \citenamefont {Zhang}}]{Zhang09}%
  \BibitemOpen
  \bibfield  {author} {\bibinfo {author} {\bibfnamefont {S.}~\bibnamefont
  {Zhang}}, \bibinfo {author} {\bibfnamefont {Y.~S.}\ \bibnamefont {Park}},
  \bibinfo {author} {\bibfnamefont {J.}~\bibnamefont {Li}}, \bibinfo {author}
  {\bibfnamefont {X.}~\bibnamefont {Lu}}, \bibinfo {author} {\bibfnamefont
  {W.}~\bibnamefont {Zhang}}, , \ and\ \bibinfo {author} {\bibfnamefont
  {X.}~\bibnamefont {Zhang}},\ }\href@noop {} {\bibfield  {journal} {\bibinfo
  {journal} {Phys. Rev. Lett.}\ }\textbf {\bibinfo {volume} {102}},\ \bibinfo
  {pages} {023901} (\bibinfo {year} {2009})}\BibitemShut {NoStop}%
\bibitem [{\citenamefont {Atre}\ \emph {et~al.}(2013)\citenamefont {Atre},
  \citenamefont {G.-Etxarri}, \citenamefont {Alaeian},\ and\ \citenamefont
  {Dionne}}]{Atre13}%
  \BibitemOpen
  \bibfield  {author} {\bibinfo {author} {\bibfnamefont {A.~C.}\ \bibnamefont
  {Atre}}, \bibinfo {author} {\bibfnamefont {A.}~\bibnamefont {G.-Etxarri}},
  \bibinfo {author} {\bibfnamefont {H.}~\bibnamefont {Alaeian}}, \ and\
  \bibinfo {author} {\bibfnamefont {J.~A.}\ \bibnamefont {Dionne}},\
  }\href@noop {} {\bibfield  {journal} {\bibinfo  {journal} {Adv. Optical
  Mater}\ }\textbf {\bibinfo {volume} {1}},\ \bibinfo {pages} {327} (\bibinfo
  {year} {2013})}\BibitemShut {NoStop}%
\bibitem [{\citenamefont {Xu}\ \emph {et~al.}(2013)\citenamefont {Xu},
  \citenamefont {Agrawal}, \citenamefont {Abashin}, \citenamefont {Chau},\ and\
  \citenamefont {Lezec}}]{Xu13}%
  \BibitemOpen
  \bibfield  {author} {\bibinfo {author} {\bibfnamefont {T.}~\bibnamefont
  {Xu}}, \bibinfo {author} {\bibfnamefont {A.}~\bibnamefont {Agrawal}},
  \bibinfo {author} {\bibfnamefont {M.}~\bibnamefont {Abashin}}, \bibinfo
  {author} {\bibfnamefont {K.~J.}\ \bibnamefont {Chau}}, \ and\ \bibinfo
  {author} {\bibfnamefont {H.~J.}\ \bibnamefont {Lezec}},\ }\href@noop {}
  {\bibfield  {journal} {\bibinfo  {journal} {Nature}\ }\textbf {\bibinfo
  {volume} {497}},\ \bibinfo {pages} {470} (\bibinfo {year}
  {2013})}\BibitemShut {NoStop}%
\bibitem [{\citenamefont {Shalaev}(2007)}]{Shalaev07}%
  \BibitemOpen
  \bibfield  {author} {\bibinfo {author} {\bibfnamefont {V.~M.}\ \bibnamefont
  {Shalaev}},\ }\href@noop {} {\bibfield  {journal} {\bibinfo  {journal} {Nat.
  Photonics}\ }\textbf {\bibinfo {volume} {1}},\ \bibinfo {pages} {41}
  (\bibinfo {year} {2007})}\BibitemShut {NoStop}%
\bibitem [{\citenamefont {Soukoulis}\ and\ \citenamefont
  {Wegener}(2011)}]{Soukoulis11}%
  \BibitemOpen
  \bibfield  {author} {\bibinfo {author} {\bibfnamefont {C.~M.}\ \bibnamefont
  {Soukoulis}}\ and\ \bibinfo {author} {\bibfnamefont {M.}~\bibnamefont
  {Wegener}},\ }\href@noop {} {\bibfield  {journal} {\bibinfo  {journal} {Nat.
  Photonics}\ }\textbf {\bibinfo {volume} {5}},\ \bibinfo {pages} {523}
  (\bibinfo {year} {2011})}\BibitemShut {NoStop}%
\bibitem [{\citenamefont {Lezec}\ \emph {et~al.}(2007)\citenamefont {Lezec},
  \citenamefont {Dionne},\ and\ \citenamefont {Atwater}}]{Lezec07}%
  \BibitemOpen
  \bibfield  {author} {\bibinfo {author} {\bibfnamefont {H.~J.}\ \bibnamefont
  {Lezec}}, \bibinfo {author} {\bibfnamefont {J.~A.}\ \bibnamefont {Dionne}}, \
  and\ \bibinfo {author} {\bibfnamefont {H.~A.}\ \bibnamefont {Atwater}},\
  }\href@noop {} {\bibfield  {journal} {\bibinfo  {journal} {Science}\ }\textbf
  {\bibinfo {volume} {316}},\ \bibinfo {pages} {430} (\bibinfo {year}
  {2007})}\BibitemShut {NoStop}%
\bibitem [{\citenamefont {Benisty}\ \emph {et~al.}(2011)\citenamefont
  {Benisty}, \citenamefont {Degiron}, \citenamefont {Lupu}, \citenamefont
  {Lustrac}, \citenamefont {Chénais}, \citenamefont {Forget}, \citenamefont
  {Besbes}, \citenamefont {Barbillon}, \citenamefont {Bruyant}, \citenamefont
  {Blaize},\ and\ \citenamefont {Lérondel}}]{Benisty11}%
  \BibitemOpen
  \bibfield  {author} {\bibinfo {author} {\bibfnamefont {H.}~\bibnamefont
  {Benisty}}, \bibinfo {author} {\bibfnamefont {A.}~\bibnamefont {Degiron}},
  \bibinfo {author} {\bibfnamefont {A.}~\bibnamefont {Lupu}}, \bibinfo {author}
  {\bibfnamefont {A.~D.}\ \bibnamefont {Lustrac}}, \bibinfo {author}
  {\bibfnamefont {S.}~\bibnamefont {Chénais}}, \bibinfo {author}
  {\bibfnamefont {S.}~\bibnamefont {Forget}}, \bibinfo {author} {\bibfnamefont
  {M.}~\bibnamefont {Besbes}}, \bibinfo {author} {\bibfnamefont
  {G.}~\bibnamefont {Barbillon}}, \bibinfo {author} {\bibfnamefont
  {A.}~\bibnamefont {Bruyant}}, \bibinfo {author} {\bibfnamefont
  {S.}~\bibnamefont {Blaize}}, \ and\ \bibinfo {author} {\bibfnamefont
  {G.}~\bibnamefont {Lérondel}},\ }\href@noop {} {\bibfield  {journal}
  {\bibinfo  {journal} {Opt. Express}\ }\textbf {\bibinfo {volume} {19}},\
  \bibinfo {pages} {18004} (\bibinfo {year} {2011})}\BibitemShut {NoStop}%
\bibitem [{\citenamefont {Guo}\ \emph {et~al.}(2009)\citenamefont {Guo},
  \citenamefont {Salamo}, \citenamefont {Duchesne}, \citenamefont {Morandotti},
  \citenamefont {Volatier-Ravat}, \citenamefont {Aimez}, \citenamefont
  {Siviloglou},\ and\ \citenamefont {Christodoulides}}]{Guo09}%
  \BibitemOpen
  \bibfield  {author} {\bibinfo {author} {\bibfnamefont {A.}~\bibnamefont
  {Guo}}, \bibinfo {author} {\bibfnamefont {G.}~\bibnamefont {Salamo}},
  \bibinfo {author} {\bibfnamefont {D.}~\bibnamefont {Duchesne}}, \bibinfo
  {author} {\bibfnamefont {R.}~\bibnamefont {Morandotti}}, \bibinfo {author}
  {\bibfnamefont {M.}~\bibnamefont {Volatier-Ravat}}, \bibinfo {author}
  {\bibfnamefont {V.}~\bibnamefont {Aimez}}, \bibinfo {author} {\bibfnamefont
  {G.}~\bibnamefont {Siviloglou}}, \ and\ \bibinfo {author} {\bibfnamefont
  {D.}~\bibnamefont {Christodoulides}},\ }\href@noop {} {\bibfield  {journal}
  {\bibinfo  {journal} {Phys. Rev. Lett.}\ }\textbf {\bibinfo {volume} {103}},\
  \bibinfo {pages} {093902} (\bibinfo {year} {2009})}\BibitemShut {NoStop}%
\bibitem [{\citenamefont {Ramezani}\ \emph {et~al.}(2012)\citenamefont
  {Ramezani}, \citenamefont {Christodoulides}, \citenamefont {Kovanis},
  \citenamefont {Vitebskiy},\ and\ \citenamefont {Kottos}}]{Ramezani12}%
  \BibitemOpen
  \bibfield  {author} {\bibinfo {author} {\bibfnamefont {H.}~\bibnamefont
  {Ramezani}}, \bibinfo {author} {\bibfnamefont {D.~N.}\ \bibnamefont
  {Christodoulides}}, \bibinfo {author} {\bibfnamefont {V.}~\bibnamefont
  {Kovanis}}, \bibinfo {author} {\bibfnamefont {I.}~\bibnamefont {Vitebskiy}},
  \ and\ \bibinfo {author} {\bibfnamefont {T.}~\bibnamefont {Kottos}},\
  }\href@noop {} {\bibfield  {journal} {\bibinfo  {journal} {Phys. Rev. Lett.}\
  }\textbf {\bibinfo {volume} {109}},\ \bibinfo {pages} {033902} (\bibinfo
  {year} {2012})}\BibitemShut {NoStop}%
\bibitem [{\citenamefont {Makris}\ \emph {et~al.}(2010)\citenamefont {Makris},
  \citenamefont {El-Ganainy},\ and\ \citenamefont {Christodoulis}}]{Makris10}%
  \BibitemOpen
  \bibfield  {author} {\bibinfo {author} {\bibfnamefont {K.}~\bibnamefont
  {Makris}}, \bibinfo {author} {\bibfnamefont {R.}~\bibnamefont {El-Ganainy}},
  \ and\ \bibinfo {author} {\bibfnamefont {D.}~\bibnamefont {Christodoulis}},\
  }\href@noop {} {\bibfield  {journal} {\bibinfo  {journal} {Phys. Rev. A}\
  }\textbf {\bibinfo {volume} {81}},\ \bibinfo {pages} {063807} (\bibinfo
  {year} {2010})}\BibitemShut {NoStop}%
\bibitem [{\citenamefont {Zheng}\ \emph {et~al.}(2010)\citenamefont {Zheng},
  \citenamefont {Christodoulides}, \citenamefont {Fleischmann},\ and\
  \citenamefont {Kottos}}]{Zheng10}%
  \BibitemOpen
  \bibfield  {author} {\bibinfo {author} {\bibfnamefont {M.~C.}\ \bibnamefont
  {Zheng}}, \bibinfo {author} {\bibfnamefont {D.~N.}\ \bibnamefont
  {Christodoulides}}, \bibinfo {author} {\bibfnamefont {R.}~\bibnamefont
  {Fleischmann}}, \ and\ \bibinfo {author} {\bibfnamefont {T.}~\bibnamefont
  {Kottos}},\ }\href@noop {} {\bibfield  {journal} {\bibinfo  {journal} {Phys.
  Rev. A}\ }\textbf {\bibinfo {volume} {82}},\ \bibinfo {pages} {010103}
  (\bibinfo {year} {2010})}\BibitemShut {NoStop}%
\bibitem [{\citenamefont {Longhi}(2011)}]{Longhi11}%
  \BibitemOpen
  \bibfield  {author} {\bibinfo {author} {\bibfnamefont {S.}~\bibnamefont
  {Longhi}},\ }\href@noop {} {\bibfield  {journal} {\bibinfo  {journal} {J.
  Phys. A: Math Theor}\ }\textbf {\bibinfo {volume} {44}},\ \bibinfo {pages}
  {485302} (\bibinfo {year} {2011})}\BibitemShut {NoStop}%
\bibitem [{\citenamefont {Rüter}\ \emph {et~al.}(2010)\citenamefont {Rüter},
  \citenamefont {Makris}, \citenamefont {El-Ganainy}, \citenamefont
  {Christodoulides}, \citenamefont {Segev},\ and\ \citenamefont
  {Kip}}]{Ruter10}%
  \BibitemOpen
  \bibfield  {author} {\bibinfo {author} {\bibfnamefont {C.~E.}\ \bibnamefont
  {Rüter}}, \bibinfo {author} {\bibfnamefont {K.~G.}\ \bibnamefont {Makris}},
  \bibinfo {author} {\bibfnamefont {R.}~\bibnamefont {El-Ganainy}}, \bibinfo
  {author} {\bibfnamefont {D.~N.}\ \bibnamefont {Christodoulides}}, \bibinfo
  {author} {\bibfnamefont {M.}~\bibnamefont {Segev}}, \ and\ \bibinfo {author}
  {\bibfnamefont {D.}~\bibnamefont {Kip}},\ }\href@noop {} {\bibfield
  {journal} {\bibinfo  {journal} {Nat. Phys.}\ }\textbf {\bibinfo {volume}
  {6}},\ \bibinfo {pages} {192} (\bibinfo {year} {2010})}\BibitemShut {NoStop}%
\bibitem [{\citenamefont {Lin}\ \emph {et~al.}(2011)\citenamefont {Lin},
  \citenamefont {Ramezani}, \citenamefont {Eichelkraut}, \citenamefont
  {Kottos}, \citenamefont {Cao},\ and\ \citenamefont
  {Christodoulides}}]{Lin11}%
  \BibitemOpen
  \bibfield  {author} {\bibinfo {author} {\bibfnamefont {Z.}~\bibnamefont
  {Lin}}, \bibinfo {author} {\bibfnamefont {H.}~\bibnamefont {Ramezani}},
  \bibinfo {author} {\bibfnamefont {T.}~\bibnamefont {Eichelkraut}}, \bibinfo
  {author} {\bibfnamefont {T.}~\bibnamefont {Kottos}}, \bibinfo {author}
  {\bibfnamefont {H.}~\bibnamefont {Cao}}, \ and\ \bibinfo {author}
  {\bibfnamefont {D.~N.}\ \bibnamefont {Christodoulides}},\ }\href@noop {}
  {\bibfield  {journal} {\bibinfo  {journal} {Phys. Rev. Lett.}\ }\textbf
  {\bibinfo {volume} {106}},\ \bibinfo {pages} {213901} (\bibinfo {year}
  {2011})}\BibitemShut {NoStop}%
\bibitem [{\citenamefont {Mostafazadeh}(2013)}]{Mostafazadeh13}%
  \BibitemOpen
  \bibfield  {author} {\bibinfo {author} {\bibfnamefont {A.}~\bibnamefont
  {Mostafazadeh}},\ }\href@noop {} {\bibfield  {journal} {\bibinfo  {journal}
  {Phys. Rev. A}\ }\textbf {\bibinfo {volume} {87}},\ \bibinfo {pages} {012103}
  (\bibinfo {year} {2013})}\BibitemShut {NoStop}%
\bibitem [{\citenamefont {Feng}\ \emph {et~al.}(2013)\citenamefont {Feng},
  \citenamefont {Xu}, \citenamefont {Fegadolli}, \citenamefont {Lu},
  \citenamefont {Oliveira}, \citenamefont {Almeida}, \citenamefont {Chen},\
  and\ \citenamefont {Scherer}}]{Feng13}%
  \BibitemOpen
  \bibfield  {author} {\bibinfo {author} {\bibfnamefont {L.}~\bibnamefont
  {Feng}}, \bibinfo {author} {\bibfnamefont {Y.~L.}\ \bibnamefont {Xu}},
  \bibinfo {author} {\bibfnamefont {W.~S.}\ \bibnamefont {Fegadolli}}, \bibinfo
  {author} {\bibfnamefont {M.~H.}\ \bibnamefont {Lu}}, \bibinfo {author}
  {\bibfnamefont {J.~E.}\ \bibnamefont {Oliveira}}, \bibinfo {author}
  {\bibfnamefont {V.~R.}\ \bibnamefont {Almeida}}, \bibinfo {author}
  {\bibfnamefont {Y.~F.}\ \bibnamefont {Chen}}, \ and\ \bibinfo {author}
  {\bibfnamefont {A.}~\bibnamefont {Scherer}},\ }\href@noop {} {\bibfield
  {journal} {\bibinfo  {journal} {Nat. Mater}\ }\textbf {\bibinfo {volume}
  {12}},\ \bibinfo {pages} {108} (\bibinfo {year} {2013})}\BibitemShut
  {NoStop}%
\bibitem [{\citenamefont {Makris}\ \emph {et~al.}(2008)\citenamefont {Makris},
  \citenamefont {El-Ganainy}, \citenamefont {Christodoulides},\ and\
  \citenamefont {Musslimani}}]{Makris08}%
  \BibitemOpen
  \bibfield  {author} {\bibinfo {author} {\bibfnamefont {K.~G.}\ \bibnamefont
  {Makris}}, \bibinfo {author} {\bibfnamefont {R.}~\bibnamefont {El-Ganainy}},
  \bibinfo {author} {\bibfnamefont {D.~N.}\ \bibnamefont {Christodoulides}}, \
  and\ \bibinfo {author} {\bibfnamefont {Z.~H.}\ \bibnamefont {Musslimani}},\
  }\href@noop {} {\bibfield  {journal} {\bibinfo  {journal} {Phys. Rev. Lett.}\
  }\textbf {\bibinfo {volume} {100}},\ \bibinfo {pages} {103904} (\bibinfo
  {year} {2008})}\BibitemShut {NoStop}%
\bibitem [{\citenamefont {G.Castaldi}\ \emph {et~al.}(2013)\citenamefont
  {G.Castaldi}, \citenamefont {Savoia}, \citenamefont {Galdi}, \citenamefont
  {Alu},\ and\ \citenamefont {Engheta}}]{Castald13}%
  \BibitemOpen
  \bibfield  {author} {\bibinfo {author} {\bibnamefont {G.Castaldi}}, \bibinfo
  {author} {\bibfnamefont {S.}~\bibnamefont {Savoia}}, \bibinfo {author}
  {\bibfnamefont {V.}~\bibnamefont {Galdi}}, \bibinfo {author} {\bibfnamefont
  {A.}~\bibnamefont {Alu}}, \ and\ \bibinfo {author} {\bibfnamefont
  {N.}~\bibnamefont {Engheta}},\ }\href@noop {} {\bibfield  {journal} {\bibinfo
   {journal} {Phys. Rev. Lett.}\ }\textbf {\bibinfo {volume} {110}},\ \bibinfo
  {pages} {173901} (\bibinfo {year} {2013})}\BibitemShut {NoStop}%
\bibitem [{\citenamefont {Musslimani}\ \emph {et~al.}(2008)\citenamefont
  {Musslimani}, \citenamefont {Makris}, \citenamefont {El-Ganainy},\ and\
  \citenamefont {Christodoulides}}]{Musslimani08}%
  \BibitemOpen
  \bibfield  {author} {\bibinfo {author} {\bibfnamefont {Z.~H.}\ \bibnamefont
  {Musslimani}}, \bibinfo {author} {\bibfnamefont {K.~G.}\ \bibnamefont
  {Makris}}, \bibinfo {author} {\bibfnamefont {R.}~\bibnamefont {El-Ganainy}},
  \ and\ \bibinfo {author} {\bibfnamefont {D.~N.}\ \bibnamefont
  {Christodoulides}},\ }\href@noop {} {\bibfield  {journal} {\bibinfo
  {journal} {Phys. Rev. Lett.}\ }\textbf {\bibinfo {volume} {100}},\ \bibinfo
  {pages} {030402} (\bibinfo {year} {2008})}\BibitemShut {NoStop}%
\bibitem [{\citenamefont {Lazarides}\ and\ \citenamefont
  {Tsironis}(2013)}]{Lazarides13}%
  \BibitemOpen
  \bibfield  {author} {\bibinfo {author} {\bibfnamefont {N.}~\bibnamefont
  {Lazarides}}\ and\ \bibinfo {author} {\bibfnamefont {G.}~\bibnamefont
  {Tsironis}},\ }\href@noop {} {\bibfield  {journal} {\bibinfo  {journal}
  {Phys. Rev. Lett.}\ }\textbf {\bibinfo {volume} {110}},\ \bibinfo {pages}
  {053901} (\bibinfo {year} {2013})}\BibitemShut {NoStop}%
\bibitem [{\citenamefont {Li}\ \emph {et~al.}(2012)\citenamefont {Li},
  \citenamefont {Huang}, \citenamefont {Liu},\ and\ \citenamefont
  {Dong}}]{Li12}%
  \BibitemOpen
  \bibfield  {author} {\bibinfo {author} {\bibfnamefont {C.}~\bibnamefont
  {Li}}, \bibinfo {author} {\bibfnamefont {C.}~\bibnamefont {Huang}}, \bibinfo
  {author} {\bibfnamefont {H.}~\bibnamefont {Liu}}, \ and\ \bibinfo {author}
  {\bibfnamefont {L.}~\bibnamefont {Dong}},\ }\href@noop {} {\bibfield
  {journal} {\bibinfo  {journal} {Opt. Lett.}\ }\textbf {\bibinfo {volume}
  {37}},\ \bibinfo {pages} {4543} (\bibinfo {year} {2012})}\BibitemShut
  {NoStop}%
\bibitem [{\citenamefont {He}\ \emph {et~al.}(2011)\citenamefont {He},
  \citenamefont {Zhu}, \citenamefont {Mihalache}, \citenamefont {Liu},\ and\
  \citenamefont {Chen}}]{He11}%
  \BibitemOpen
  \bibfield  {author} {\bibinfo {author} {\bibfnamefont {Y.}~\bibnamefont
  {He}}, \bibinfo {author} {\bibfnamefont {X.}~\bibnamefont {Zhu}}, \bibinfo
  {author} {\bibfnamefont {D.}~\bibnamefont {Mihalache}}, \bibinfo {author}
  {\bibfnamefont {J.}~\bibnamefont {Liu}}, \ and\ \bibinfo {author}
  {\bibfnamefont {Z.}~\bibnamefont {Chen}},\ }\href@noop {} {\bibfield
  {journal} {\bibinfo  {journal} {Phys. Rev. A}\ }\textbf {\bibinfo {volume}
  {85}},\ \bibinfo {pages} {013831} (\bibinfo {year} {2011})}\BibitemShut
  {NoStop}%
\bibitem [{\citenamefont {Dmitriev}\ \emph {et~al.}(2010)\citenamefont
  {Dmitriev}, \citenamefont {Sukhorukov},\ and\ \citenamefont
  {Kivshar}}]{Dmitriev10}%
  \BibitemOpen
  \bibfield  {author} {\bibinfo {author} {\bibfnamefont {S.~V.}\ \bibnamefont
  {Dmitriev}}, \bibinfo {author} {\bibfnamefont {A.~A.}\ \bibnamefont
  {Sukhorukov}}, \ and\ \bibinfo {author} {\bibfnamefont {Y.~S.}\ \bibnamefont
  {Kivshar}},\ }\href@noop {} {\bibfield  {journal} {\bibinfo  {journal} {Opt.
  Lett.}\ }\textbf {\bibinfo {volume} {35}},\ \bibinfo {pages} {2976} (\bibinfo
  {year} {2010})}\BibitemShut {NoStop}%
\bibitem [{\citenamefont {Chong}\ \emph {et~al.}(2011)\citenamefont {Chong},
  \citenamefont {Ge},\ and\ \citenamefont {Stone}}]{Chong11}%
  \BibitemOpen
  \bibfield  {author} {\bibinfo {author} {\bibfnamefont {Y.~D.}\ \bibnamefont
  {Chong}}, \bibinfo {author} {\bibfnamefont {L.}~\bibnamefont {Ge}}, \ and\
  \bibinfo {author} {\bibfnamefont {A.~D.}\ \bibnamefont {Stone}},\ }\href@noop
  {} {\bibfield  {journal} {\bibinfo  {journal} {Phys. Rev. Lett.}\ }\textbf
  {\bibinfo {volume} {106}},\ \bibinfo {pages} {093902} (\bibinfo {year}
  {2011})}\BibitemShut {NoStop}%
\bibitem [{\citenamefont {Longhi}(2010)}]{Longhi10}%
  \BibitemOpen
  \bibfield  {author} {\bibinfo {author} {\bibfnamefont {S.}~\bibnamefont
  {Longhi}},\ }\href@noop {} {\bibfield  {journal} {\bibinfo  {journal} {Phys.
  Rev. A}\ }\textbf {\bibinfo {volume} {82}},\ \bibinfo {pages} {031801}
  (\bibinfo {year} {2010})}\BibitemShut {NoStop}%
\bibitem [{\citenamefont {Bender}\ \emph {et~al.}(1999)\citenamefont {Bender},
  \citenamefont {Dune},\ and\ \citenamefont {Meisinger}}]{Bender99}%
  \BibitemOpen
  \bibfield  {author} {\bibinfo {author} {\bibfnamefont {C.~M.}\ \bibnamefont
  {Bender}}, \bibinfo {author} {\bibfnamefont {G.~V.}\ \bibnamefont {Dune}}, \
  and\ \bibinfo {author} {\bibfnamefont {P.~N.}\ \bibnamefont {Meisinger}},\
  }\href@noop {} {\bibfield  {journal} {\bibinfo  {journal} {Phys. Lett. A}\
  }\textbf {\bibinfo {volume} {252}},\ \bibinfo {pages} {272} (\bibinfo {year}
  {1999})}\BibitemShut {NoStop}%
\bibitem [{\citenamefont {Bender}(2007)}]{Bender07}%
  \BibitemOpen
  \bibfield  {author} {\bibinfo {author} {\bibfnamefont {C.~M.}\ \bibnamefont
  {Bender}},\ }\href@noop {} {\bibfield  {journal} {\bibinfo  {journal} {Rep.
  Prog. Phys.}\ }\textbf {\bibinfo {volume} {70}},\ \bibinfo {pages} {947}
  (\bibinfo {year} {2007})}\BibitemShut {NoStop}%
\bibitem [{\citenamefont {Ahmed}(2001)}]{Ahmed01}%
  \BibitemOpen
  \bibfield  {author} {\bibinfo {author} {\bibfnamefont {Z.}~\bibnamefont
  {Ahmed}},\ }\href@noop {} {\bibfield  {journal} {\bibinfo  {journal} {Phys.
  Lett. A}\ }\textbf {\bibinfo {volume} {282}},\ \bibinfo {pages} {343}
  (\bibinfo {year} {2001})}\BibitemShut {NoStop}%
\bibitem [{\citenamefont {Bender}\ and\ \citenamefont
  {Boettcher}(1998)}]{Bender98}%
  \BibitemOpen
  \bibfield  {author} {\bibinfo {author} {\bibfnamefont {C.~M.}\ \bibnamefont
  {Bender}}\ and\ \bibinfo {author} {\bibfnamefont {S.}~\bibnamefont
  {Boettcher}},\ }\href@noop {} {\bibfield  {journal} {\bibinfo  {journal}
  {Phys. Rev. Lett.}\ }\textbf {\bibinfo {volume} {80}},\ \bibinfo {pages}
  {5243} (\bibinfo {year} {1998})}\BibitemShut {NoStop}%
\bibitem [{\citenamefont {Mostafazadeh}(2002)}]{Mostafazadeh02}%
  \BibitemOpen
  \bibfield  {author} {\bibinfo {author} {\bibfnamefont {A.}~\bibnamefont
  {Mostafazadeh}},\ }\href@noop {} {\bibfield  {journal} {\bibinfo  {journal}
  {J. Math. Phys.}\ }\textbf {\bibinfo {volume} {43}},\ \bibinfo {pages} {2814}
  (\bibinfo {year} {2002})}\BibitemShut {NoStop}%
\bibitem [{\citenamefont {Verhagen}\ \emph {et~al.}(2010)\citenamefont
  {Verhagen}, \citenamefont {de~Waele}, \citenamefont {Kuipers},\ and\
  \citenamefont {Polman}}]{Verhagen10}%
  \BibitemOpen
  \bibfield  {author} {\bibinfo {author} {\bibfnamefont {E.}~\bibnamefont
  {Verhagen}}, \bibinfo {author} {\bibfnamefont {R.}~\bibnamefont {de~Waele}},
  \bibinfo {author} {\bibfnamefont {L.}~\bibnamefont {Kuipers}}, \ and\
  \bibinfo {author} {\bibfnamefont {A.}~\bibnamefont {Polman}},\ }\href@noop {}
  {\bibfield  {journal} {\bibinfo  {journal} {Phys. Rev. Lett.}\ }\textbf
  {\bibinfo {volume} {105}},\ \bibinfo {pages} {223901} (\bibinfo {year}
  {2010})}\BibitemShut {NoStop}%
\bibitem [{Note1()}]{Note1}%
  \BibitemOpen
  \bibinfo {note} {Here we assume a lossless Drude model, but the results are
  generalizable to a realistic case including loss. See for example reference
  ~\cite {Benisty11}.}\BibitemShut {Stop}%
\bibitem [{\citenamefont {Russell}\ \emph {et~al.}(1995)\citenamefont
  {Russell}, \citenamefont {Birks},\ and\ \citenamefont
  {Lloyd-Lucas}}]{Russell95}%
  \BibitemOpen
  \bibfield  {author} {\bibinfo {author} {\bibfnamefont {P.~S.~J.}\
  \bibnamefont {Russell}}, \bibinfo {author} {\bibfnamefont {T.~A.}\
  \bibnamefont {Birks}}, \ and\ \bibinfo {author} {\bibfnamefont {F.~D.}\
  \bibnamefont {Lloyd-Lucas}},\ }\href@noop {} {\bibfield  {journal} {\bibinfo
  {journal} {Confined Electrons and Photons}\ } (\bibinfo {year}
  {1995})}\BibitemShut {NoStop}%
\bibitem [{\citenamefont {Ni}\ \emph {et~al.}(2011)\citenamefont {Ni},
  \citenamefont {Naik}, \citenamefont {Kildishev}, \citenamefont {Barnakov},
  \citenamefont {Boltasseva},\ and\ \citenamefont {Shalaev}}]{Ni11}%
  \BibitemOpen
  \bibfield  {author} {\bibinfo {author} {\bibfnamefont {X.}~\bibnamefont
  {Ni}}, \bibinfo {author} {\bibfnamefont {G.}~\bibnamefont {Naik}}, \bibinfo
  {author} {\bibfnamefont {A.}~\bibnamefont {Kildishev}}, \bibinfo {author}
  {\bibfnamefont {Y.}~\bibnamefont {Barnakov}}, \bibinfo {author}
  {\bibfnamefont {A.}~\bibnamefont {Boltasseva}}, \ and\ \bibinfo {author}
  {\bibfnamefont {V.}~\bibnamefont {Shalaev}},\ }\href@noop {} {\bibfield
  {journal} {\bibinfo  {journal} {Appl. Phys. B}\ }\textbf {\bibinfo {volume}
  {103}},\ \bibinfo {pages} {553} (\bibinfo {year} {2011})}\BibitemShut
  {NoStop}%
\bibitem [{\citenamefont {Jacob}\ \emph {et~al.}(2012)\citenamefont {Jacob},
  \citenamefont {Smolyaninov},\ and\ \citenamefont {Narimanov}}]{Jacob12}%
  \BibitemOpen
  \bibfield  {author} {\bibinfo {author} {\bibfnamefont {Z.}~\bibnamefont
  {Jacob}}, \bibinfo {author} {\bibfnamefont {I.~I.}\ \bibnamefont
  {Smolyaninov}}, \ and\ \bibinfo {author} {\bibfnamefont {E.~E.}\ \bibnamefont
  {Narimanov}},\ }\href@noop {} {\bibfield  {journal} {\bibinfo  {journal}
  {App. Phys. Lett.}\ }\textbf {\bibinfo {volume} {100}},\ \bibinfo {pages}
  {181105} (\bibinfo {year} {2012})}\BibitemShut {NoStop}%
\bibitem [{\citenamefont {Webb}\ and\ \citenamefont {Yang}(2006)}]{Webb06}%
  \BibitemOpen
  \bibfield  {author} {\bibinfo {author} {\bibfnamefont {K.~J.}\ \bibnamefont
  {Webb}}\ and\ \bibinfo {author} {\bibfnamefont {M.}~\bibnamefont {Yang}},\
  }\href@noop {} {\bibfield  {journal} {\bibinfo  {journal} {Opt. Lett.}\
  }\textbf {\bibinfo {volume} {31}},\ \bibinfo {pages} {2130} (\bibinfo {year}
  {2006})}\BibitemShut {NoStop}%
\bibitem [{\citenamefont {Kim}\ \emph {et~al.}(2012)\citenamefont {Kim},
  \citenamefont {Drachev}, \citenamefont {Jacob}, \citenamefont {Naik},
  \citenamefont {Boltasseva}, \citenamefont {Narimanov},\ and\ \citenamefont
  {Shalaev}}]{Kim12}%
  \BibitemOpen
  \bibfield  {author} {\bibinfo {author} {\bibfnamefont {J.}~\bibnamefont
  {Kim}}, \bibinfo {author} {\bibfnamefont {V.}~\bibnamefont {Drachev}},
  \bibinfo {author} {\bibfnamefont {Z.}~\bibnamefont {Jacob}}, \bibinfo
  {author} {\bibfnamefont {G.}~\bibnamefont {Naik}}, \bibinfo {author}
  {\bibfnamefont {A.}~\bibnamefont {Boltasseva}}, \bibinfo {author}
  {\bibfnamefont {E.}~\bibnamefont {Narimanov}}, \ and\ \bibinfo {author}
  {\bibfnamefont {V.}~\bibnamefont {Shalaev}},\ }\href@noop {} {\bibfield
  {journal} {\bibinfo  {journal} {Opt. Express}\ }\textbf {\bibinfo {volume}
  {20}},\ \bibinfo {pages} {8100–8116} (\bibinfo {year} {2012})}\BibitemShut
  {NoStop}%
\bibitem [{\citenamefont {Zhao}\ \emph {et~al.}(2011)\citenamefont {Zhao},
  \citenamefont {Huang},\ and\ \citenamefont {Lu}}]{Zhao11}%
  \BibitemOpen
  \bibfield  {author} {\bibinfo {author} {\bibfnamefont {W.}~\bibnamefont
  {Zhao}}, \bibinfo {author} {\bibfnamefont {X.}~\bibnamefont {Huang}}, \ and\
  \bibinfo {author} {\bibfnamefont {Z.}~\bibnamefont {Lu}},\ }\href@noop {}
  {\bibfield  {journal} {\bibinfo  {journal} {Opt. Express}\ }\textbf {\bibinfo
  {volume} {19}},\ \bibinfo {pages} {15297} (\bibinfo {year}
  {2011})}\BibitemShut {NoStop}%
\bibitem [{\citenamefont {Sapienza}\ \emph {et~al.}(2003)\citenamefont
  {Sapienza}, \citenamefont {Costantino},\ and\ \citenamefont
  {Wiersma}}]{Sapienza03}%
  \BibitemOpen
  \bibfield  {author} {\bibinfo {author} {\bibfnamefont {R.}~\bibnamefont
  {Sapienza}}, \bibinfo {author} {\bibfnamefont {P.}~\bibnamefont
  {Costantino}}, \ and\ \bibinfo {author} {\bibfnamefont {D.}~\bibnamefont
  {Wiersma}},\ }\href@noop {} {\bibfield  {journal} {\bibinfo  {journal} {Phys.
  Rev. Lett.}\ }\textbf {\bibinfo {volume} {91}},\ \bibinfo {pages} {263902}
  (\bibinfo {year} {2003})}\BibitemShut {NoStop}%
\bibitem [{\citenamefont {Ge}\ \emph {et~al.}(2012)\citenamefont {Ge},
  \citenamefont {Chong},\ and\ \citenamefont {Stone}}]{Ge12}%
  \BibitemOpen
  \bibfield  {author} {\bibinfo {author} {\bibfnamefont {L.}~\bibnamefont
  {Ge}}, \bibinfo {author} {\bibfnamefont {Y.}~\bibnamefont {Chong}}, \ and\
  \bibinfo {author} {\bibfnamefont {A.}~\bibnamefont {Stone}},\ }\href@noop {}
  {\bibfield  {journal} {\bibinfo  {journal} {Phys. Rev. A}\ }\textbf {\bibinfo
  {volume} {85}},\ \bibinfo {pages} {023802} (\bibinfo {year}
  {2012})}\BibitemShut {NoStop}%
\end{thebibliography}%

\end{document}